# Optimization of LMS Algorithm for System Identification

*Saurabh R. Prasad[1*], Bhalchandra B. Godbole[2]*
[1]Department of Electronics and Telecommunication Engineering, DKTE Society's Textile and Engineering Institute, Ichalkaranji, India
[2]Department of Electronics Engineering, KBP College of Engineering, Satara, India

* Corresponding Author:
Saurabh R Prasad
saurabhprasad21@gmail.com

**Abstract**-An adaptive filter is defined as a digital filter that has the capability of self adjusting its transfer function under the control of some optimizing algorithms. Most common optimizing algorithms are Least Mean Square (LMS) and Recursive Least Square (RLS). Although RLS algorithm perform superior to LMS algorithm, it has very high computational complexity so not useful in most of the practical scenario. So most feasible choice of the adaptive filtering algorithm is the LMS algorithm including its various variants. The LMS algorithm uses transversal FIR filter as underlying digital filter. This paper is based on implementation and optimization of LMS algorithm for the application of unknown system identification.

**Keywords**- Adaptive Filtering, LMS Algorithm, Optimization, System Identification, MATLAB

## 1. INTRODUCTION TO ADAPTIVE FILTERING

There are two types of digital filters; fixed filter and adaptive filter. A fixed filter is useful when the parameters of signal and channel are known. On the other hand, adaptive filters are useful when the dynamics of signal or channel is unpredictable and changes with time. In such applications, adaptive filters are useful which adapt to these changes to obtain the desired output. Adaptive filters possesses a wide range of applicability such as system identification, inverse system modeling, equalization, interference cancellation, acoustic and network echo cancellation, adaptive beam-forming etc [1-2]. As per the applications of adaptive filters there are four distinct types of configurations as depicted in fig.1.

### 1.1 System Identification

System identification, also popularly called as mathematical modeling is a category of adaptive filtering that finds a huge range of applications particularly in the areas of communication. Adaptive filter is used to provide the linear model which represents the best fit of the unknown plant. In this configuration, the plant and the adaptive filter are connected in parallel form and driven by the same input. The plant output is called as desired response of the system. The error signal is obtained by subtracting the adaptive filter output from the desired output. If the plant is dynamic, then adaptive filter will also be time varying or non stationary.

## 1.2 Inverse Modeling

In this configuration, adaptive filter's role is to find the inverse model that represents the best fit for the unknown plant. The unknown plant and the adaptive filter are connected in cascade. Ideally here, the adaptive filter transfer function should be the reciprocal of the plant's transfer function so that when both are connected in cascade, they will represent ideal channel. Here plant is driven by the input signal and adaptive filter is driven by the output of the plant. This configuration of adaptive filters is used in equalizers.

## 1.3 Prediction

Here the role of adaptive filter is to find the best prediction of the instantaneous value of the input signal. The present value thus acts as desired response of the input signal. Delay network is incorporated to provide time delay to the input signal. The delayed signal is applied as the input to the filter. This filter can be used for the purpose of predictive coding and spectrum analysis.

## 1.4 Interference Cancellation

In this configuration adaptive filter is used to cancel the unknown interference contained in the primary signal also called as desired signal. This configuration is also used in noise cancellation and beam-forming applications.

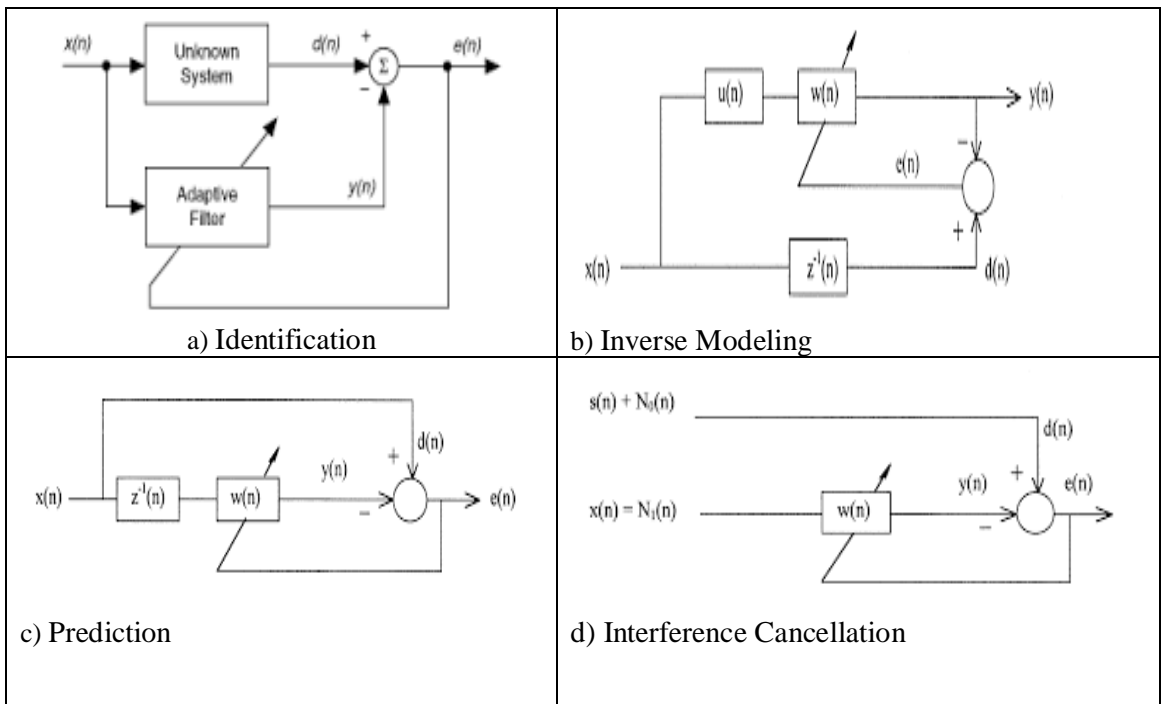

a) Identification  
b) Inverse Modeling  
c) Prediction  
d) Interference Cancellation  

Fig1. Configurations of Adaptive Filter

An adaptive filter consists of two elements, a digital filter and an adaptive algorithm. As shown in Fig. 2, the digital filter is most probably a FIR filter that performs the filtering operation to produce output in response to the input. The adaptive algorithm is used to iteratively adjust the coefficients of the digital filter in order to minimize the cost function. The most common and widely used adaptive algorithm is known as Least Mean Square (LMS) algorithm. Both adaptive and fixed digital filter need to be stable and causal. The general schematic diagram of adaptive filter and the underlying direct form transversal structure are as represented in fig.2.

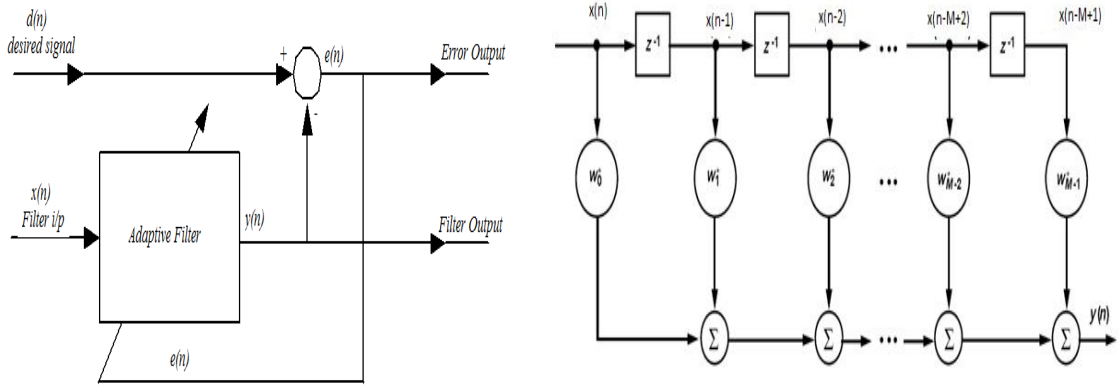

a) Adaptive Filter Schematics  b) Underlying Transversal Filter

Fig. 2. a) Adaptive Filter Schematics, b) Underlying Transversal FIR Filter Structure

### 1.5 Weiner filter

*Weiner* filter is the fundamental concept in the theory of adaptive filter. It is an optimal filter and was derived by two researchers independently, namely *Wiener* in 1942, and *Kolmogorov* in 1939 [3-4]. Actually, the filter coefficients as well as input signal both may be complex quantities. However, the practical signal of interest and the filter are mostly real quantities. For this reason and in order to simplify the derivation presented here, Weiner filter is derived for real signal and filter coefficients [5]. There are various types of optimization criteria. Wiener solution finds the filter weight vector $\mathbf{w}(n)$ so that the expected value of squared error $E[e^2(n)]$ gets minimized [6-8]. This criterion is known as Mean Square Error (MSE) criterion.

Consider following notations,

$\mathbf{x}(n)$ : $[x(n)\ x(n-1)\ \ldots\ldots x(n-N+1)]^T$, input signal vector

$\mathbf{w}(n)$: $[w(n)\ w(n-1)\ \ldots w(n-N+1)]^T$, adaptive filter weight vector

$\mu$   : step size or convergence parameter

$y(n)$ : $\mathbf{w}^T \mathbf{x}$, instantaneous value of FIR filter output

$d(n)$ : instantaneous value of the desired signal

$e(n)$ : $d(n)-y(n)$, instantaneous value of error signal

$\mathbf{w_o}$   : weight of optimal filter

$\mathbf{R}$    : $E[\mathbf{x}(n)\ \mathbf{x}^T(n)]$, Correlation matrix of $\mathbf{x}(n)$

$\mathbf{p}$: $E[d(n)\mathbf{x}(n)]$, Cross correlation vector between $\mathbf{x}(n)$ and $d(n)$

$J_n$  : $E[e^2(n)]$, Mean Square Error Cost function

$N$   : Order of the filter

The output of transversal filter is represented as,

$$y(n) = w_o x(n) + w_1 x(n-1) + \ldots + w_{N-1} x(n-N+1)$$
$$y(n) = \mathbf{w}^T \mathbf{x(n)} \quad (1)$$

MSE cost function is given by,

$$J(\mathbf{w}) = E[e^2(n)] \quad \text{(2-a)}$$
$$J(\mathbf{w}) = E\{[d(n)-y(n)]^2\}$$
$$J(\mathbf{w}) = E\{d^2(n)-2d(n)y(n)+y^2(n)\}$$

Substituting for $y(n)$ from (1) and considering the fact that expectation operator is linear,

$$J(\mathbf{w}) = E\{d^2(n)-2d(n)\mathbf{w}^T\mathbf{x}(n)+\mathbf{w}^T\mathbf{x}(n)\mathbf{x}^T(n)\mathbf{w}\}$$

As filter weight $\mathbf{w}$ is not a random variable, the cost function reduces to,

$$J(\mathbf{w}) = E\{d^2(n)\}-2\mathbf{w}^T E\{d(n)\mathbf{x}(n)\}+\mathbf{w}^T E\{\mathbf{x}(n)\mathbf{x}^T(n)\}\mathbf{w}$$

Assume desired signal $d(n)$ is zero mean. Now, the term $E\{d^2(n)\}$ is same as variance of the desired signal, and given by $\sigma_d^2$. Substituting the expression for $\mathbf{R}$ and $\mathbf{p}$, the above equation reduces to,

$$J(\mathbf{w}) = \sigma_d^2 - 2\mathbf{w}^T\mathbf{p} + \mathbf{w}^T\mathbf{R}\mathbf{w} \quad \text{(2-b)}$$

From (2-b), it is observed that the $J(\mathbf{w})$ is linear due to second term in this equation and quadratic due to third term in this equation. Thus the overall equation of cost function $J(\mathbf{w})$ is quadratic. It is well known that there exists a unique minimum point for a quadratic or convex equation, and it can be solved by taking the gradient of equation (2-b). Thus we obtain,

$$\nabla J(\mathbf{w}) = -2\mathbf{p} + 2\mathbf{R}\mathbf{w} \quad \text{(3)}$$

By equating the gradient to zero, we get the Wiener solution as given in (4),

$$\mathbf{w}_{opt} = \mathbf{R}^{-1}\mathbf{p} \quad \text{(4)}$$

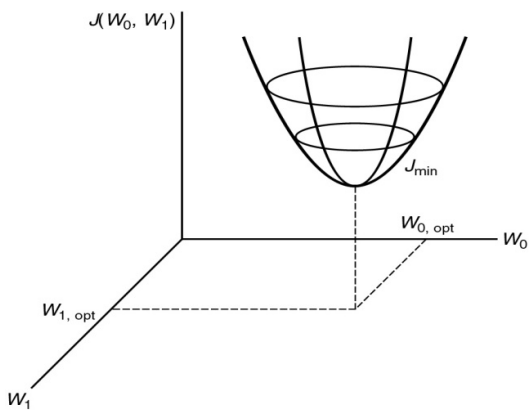

Fig.3. Optimal Weiner Solution of Adaptive Filter Weight

Optimal Weiner solution for two coefficient adaptive filter is shown in fig.3. Weiner gives the optimal solution in the sense that no other method can achieve lower value of the MSE cost function than the Weiner solution. Still it is not practically possible in real time scenario due to various reasons that can be identified from (4). Equation (4) indicates that computation of optimal solution requires correlation matrix inversion which is quite computationally intensive. The computation of optimal solution also requires correlation matrix $\mathbf{R}$ and cross correlation vector $\mathbf{p}$ those are not directly available. The above mentioned points suggest an iterative approach to calculate Wiener solution instead of direct solution using (4). A number of algorithms exist for finding the optimal solution iteratively. However selection of a particular algorithm depends upon application, rate of convergence of algorithm versus required speed, computational complexity, target hardware, memory requirement, latency, and accuracy etc. The iterative methods to find optimal solution are discussed in the following section.

*A. Newton's Algorithms*

Let $\mathbf{w}(n+1)$ be the filter weight vector at $(n+1)^{th}$ iteration. The filter weight update equation is given in (5), where $\mathbf{g}(n)$ is the direction vector.

$$\mathbf{w}(n+1) = \mathbf{w}(n) + \mathbf{g}(n) \tag{5}$$

Now we have to calculate the direction vector so as to minimize $J(\mathbf{w}(n+1))$ i.e. the cost function at next iteration.

$$J(\mathbf{w}(n+1)) = J(\mathbf{w}(n)+\mathbf{g}(n)) \tag{6}$$

The Taylor series expansion of (6) is given in (7)

$$J(\mathbf{w}+\mathbf{g}) = J(\mathbf{w}) + \sum_i g_i \frac{\partial J(\mathbf{w})}{\partial w_i} + \frac{1}{2}\sum_i \sum_j g_i g_j \frac{\partial^2 J(\mathbf{w})}{\partial w_i \partial w_j} + \ldots \tag{7}$$

Considering adaptive filter of first order, there are two filter coefficients $w_0$ and $w_1$ and above equation can be modified as,

$$J(\mathbf{w}+\mathbf{g}) = J(\mathbf{w}) + g_o \frac{\partial J(\mathbf{w})}{\partial w_o} + g_1 \frac{\partial J(\mathbf{w})}{\partial w_1} + \frac{1}{2}[g_o^2 \frac{\partial^2 J(\mathbf{w})}{\partial w_o \partial w_o} + g_o g_1 \frac{\partial^2 J(\mathbf{w})}{\partial w_o \partial w_1} + g_o g_1 \frac{\partial^2 J(\mathbf{w})}{\partial w_1 \partial w_o} + g_1^2 \frac{\partial^2 J(\mathbf{w})}{\partial w_1 \partial w_1}] \tag{8}$$

Taking partial differentiation of (8) w.r.t. $g_o$ and $g_1$ gives,

$$\frac{\partial J(\mathbf{w}+\mathbf{g})}{\partial g_o} = \frac{\partial J(\mathbf{w})}{\partial w_o} + g_o \frac{\partial^2 J(\mathbf{w})}{\partial w_o \partial w_o} + g_1 \frac{\partial J(\mathbf{w})}{\partial w_o \partial w_1} \tag{9}$$

$$\frac{\partial J(\mathbf{w}+\mathbf{g})}{\partial g_1} = \frac{\partial J(\mathbf{w})}{\partial w_1} + g_o \frac{\partial^2 J(\mathbf{w})}{\partial w_o \partial w_1} + g_1 \frac{\partial^2 J(\mathbf{w})}{\partial w_1 \partial w_1} \tag{10}$$

Rearranging (9) and (10) in matrix form,

$$\frac{\partial J(\mathbf{w}+\mathbf{g})}{\partial \mathbf{g}} = \nabla J + \begin{bmatrix} \frac{\partial^2 J(\mathbf{w})}{\partial w_o \partial w_o} & \frac{\partial^2 J(\mathbf{w})}{\partial w_o \partial w_1} \\ \frac{\partial^2 J(\mathbf{w})}{\partial w_o \partial w_1} & \frac{\partial^2 J(\mathbf{w})}{\partial w_1 \partial w_1} \end{bmatrix} \begin{bmatrix} g_0 \\ g_1 \end{bmatrix} \tag{11}$$

The matrix in (11) is known as Hessian matrix, $\mathbf{H}$. To find the direction vector such that $J(\mathbf{w}+\mathbf{g})$ is minimum, (11) is equated to zero, which gives,

$$\mathbf{g} = -\mathbf{H}^{-1} \nabla J \tag{12}$$

Substituting (12) in (5) gives the weight update equation as,

$$\mathbf{w}(n+1) = \mathbf{w}(n) - \mathbf{H}^{-1} \nabla J \tag{13}$$

Above equation (13) is known as Newton's algorithm. This equation indicates that Newton's algorithm converge in one step only if the Hessian $\mathbf{H}$ and gradient $\nabla \mathbf{J}$ are known [9-10]. But Newton's equation doesn't come for rescue as still it requires matrix inversion $\mathbf{H}^-$. So, instead of direct computation of $\mathbf{H}^{-1}$, various iterative methods are used that results into different algorithms. Here Newton's algorithm is derived for second order case, but it can be used for any order using similar analytical treatment.

*B. Steepest Descent Algorithm: an Iterative Approach*

The SDA algorithm finds the local minimum of the cost function. It performs approximation of $\mathbf{H}^{-1}$ in Newton's method presented in (13) by substituting $\mathbf{H} = 2\mathbf{I}$, and an additional parameter called as convergence rate parameter μ (>0) is introduced. Thus (13) gets converted into,

$$\mathbf{w}(n+1) = \mathbf{w}(n) - \frac{\mu}{2} \nabla \mathbf{J} \tag{14}$$

Substituting gradient from (3) into (14) gives update equation of SDA as,

$$\mathbf{w}(n+1) = \mathbf{w}(n) + \mu[\mathbf{p} - \mathbf{R}\mathbf{w}(n)] \tag{15}$$

From (15) it is observed that SDA requires presence of parameters **R** and **p,** but for most of the real time applications these are not known. To address this issue, LMS algorithm is devised.

## 2. LMS ALGORITHM

The research work on adaptive filtering can be traced back to 1950s when a number of researchers were independently working on the different applications of adaptive filters. From this study, LMS algorithm emerged as a simple yet effective algorithm for the purpose of adaptive filtering. It was devised by *Bernard Widrow,* Professor of Stanford University and his doctoral research scholar, *Ted Hoff* in 1959. LMS algorithm is known as a stochastic gradient algorithm. This means that it iterates each tap weight of transversal filter in the direction of the gradient of the squared magnitude of error with respect to the tap weight. The filter is only adapted based on the error at the current time. LMS algorithm is closely related to the concept of the stochastic approximation developed by Robbins and Monro in 1951. The primary difference between the two is LMS algorithm uses the fixed convergence rate parameter to update the tap weight of the filter whereas stochastic approximation method uses the convergence parameter that in inversely proportional to the time n or power of n. There is another stochastic gradient algorithm which is closely related to LMS. This algorithm is called as Gradient Adaptive Lattice (GAL) developed by Griffiths in 1977. The difference between the LMS and GAL is only in underlying filtering structure, LMS uses transversal structure and GAL uses lattice structure. The LMS algorithm has an inherent limitation that it can search local minima only but not global minima [8]. However, this limitation can be overcome by simultaneously initializing the search at multiple points. This algorithm is derived as follows.

The LMS algorithm is an approximate version of SDA which simply approximates **R** and **p** by replacing the expectation operator by instantaneous value. Thus,

$$\mathbf{R} = E[\mathbf{x}(n)\mathbf{x}^T(n)] \approx \mathbf{x}(n)\mathbf{x}^T(n) \tag{16}$$
$$\mathbf{p} = E[d(n)\mathbf{x}(n)] \approx d(n)\mathbf{x}(n) \tag{17}$$

Substituting (16) and (17) in Steepest Descent algorithm given in (15) gives,
$\mathbf{w}(n+1) = \mathbf{w}(n) + \mu[d(n)\mathbf{x}(n) - \mathbf{x}(n)\mathbf{x}^T(n)\mathbf{w}(n)]$
By separating out the common factor,
$\mathbf{w}(n+1) = \mathbf{w}(n) + \mu \mathbf{x}(n)[d(n) - \mathbf{x}^T(n)\mathbf{w}(n)]$
Using the relationship $y(n) = \mathbf{x}^T(n)\mathbf{w}(n)$ and $e(n) = d(n) - y(n)$, above equation can be written as
$$\mathbf{w}(n+1) = \mathbf{w}(n) + \mu \mathbf{x}(n) e(n) \tag{18}$$

This equation represents popular LMS algorithm equation. This algorithm is of O(N) and requires 2N+1 multiplication and 2N addition per iteration. For stability of LMS algorithm convergence rate parameter μ satisfies the relationship,

$$0 < \mu < \frac{2}{\lambda_{max}} \tag{19}$$

where $\lambda_{max}$ is largest eigenvalues of the correlation matrix. In real time scenario, eigenvalues of the correlation matrix are not known which requires to modify (19) as,

$$0 < \mu < \frac{2}{\|x(n)\|^2} \quad (20)$$

where $\|x(n)\|$ is called as Euclidean norm of the input vector which represents the power of the signal that is usually known or can be estimated a priori.

## 2.1 Misadjustment

The misadjustment represents how far is the iterative solution at steady state condition from the optimal solution of cost function. Thus it is represented as,

$$M = \frac{J_{ex}}{J_{min}}$$

$$M = \frac{J_{ss} - J_{min}}{J_{min}} \quad (21)$$

where $J_{ex}$ represents excess MSE, $J_{ss}$ represents steady state MSE and $J_{min}$ represents optimal MSE given by Weiner solution.

## 2.2 Robustness under H∞ criteria

Different researchers were working on the robust design issue. *Zames* in 1981 introduced $H^\infty$ or minimax criteria as robust index of performance. *Hassibi* in 1996 shown that LMS algorithm is optimal under $H^\infty$ criteria which presented theoretical evidence for the robust performance of LMS algorithm.

## 2.3 Various Forms of LMS Algorithm

LMS algorithm has so far attracted various researchers working in the area of adaptive signal processing. It has remained as prime focus of research in adaptive filtering, and produced so many modified forms. Few important variants include Normalized LMS algorithm, Variable step size LMS (VSS-LMS), Variable length LMS (VL-LMS), Scrambled LMS (SCLMS), Block LMS, Signed LMS etc. In following section, TDLMS, and VL-LMS are described.

### 2.3.1 Transform Domain LMS (TDLMS)

We know that the performance of the LMS algorithm degrades if the input signal is highly correlated. In other words if there is higher Eigen value spread of the auto-correlation matrix **R** then the performance is poor and vice-versa. To overcome this problem, Transform Domain LMS algorithms are used which basically performs in order to de-correlate the input signal and improve the performance. The block diagram representing implementation of TD-LMS is given in Fig.4.

Some important points of this algorithm are as mentioned below.
a) As seen in figure, input signal u(n) and desired signal d(n) are transformed in the frequency domain by FFT.
b) Then the filtering is done on transformed signal U to produce the transformed output signal Y. The inverse transform of Y is obtained to get the output y(n) in time domain.
c) The input signal, weight vector, and output are as given below.
$u_i(n) = u(iL+n)$, n=0,1,..L-1, i=0, 1, 2…
$d_i(n) = d(iL+n)$, n=0,1,..L-1, i=0, 1, 2…
$Y_i(k) = \mathbf{W}_i(k)\,\mathbf{U}_i(k)$ …k = 0,1,…L-1;

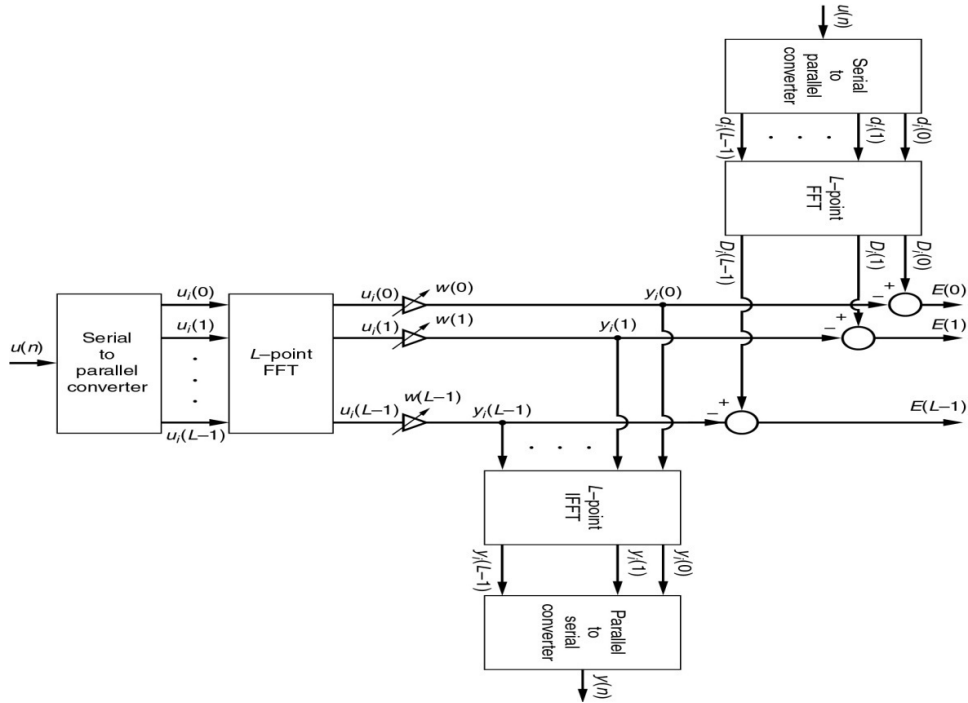

Fig.4. TDLMS Algorithm

The update equation is given as,

$$\mathbf{W}_{i+1}(k) = \mathbf{W}_i(k) + \mu_f \mathbf{E}_i(k)\mathbf{U}_k^*(k); \text{ where } Ei(k) = Di(k) - Yi(k)$$

$$\mathbf{W}_i = [\mathbf{W}_i(0)\ \mathbf{W}_i(1)\ \mathbf{W}_i(2)\ ...\ \mathbf{W}_i(L-1)]^T$$

$$\mathbf{E}_i = [\mathbf{E}_i(0)\ \mathbf{E}(1)\ \mathbf{E}(2)\ ...\ \mathbf{E}_i(L-1)]^T$$

$$\mathbf{U}_i = \begin{bmatrix} U_i(0) & 0 & 0 & ... & 0 \\ 0 & U_i(1) & 0 & ... & 0 \\ 0 & 0 & 0 & ... & 0 \\ ... & ... & ... & ... & ... \\ 0 & 0 & 0 & ... & U_i(L-1) \end{bmatrix}$$

d) In time domain LMS algorithm, the weight is updated for every sample, but in transform domain LMS the weight vector is updated only once for every block of data. This reduces computational complexity and increases the rate of convergence.

**2.3.2 Variable Length LMS Algorithm**

In the applications of system identification, the adaptive filter produces good accuracy if the number of filter coefficients is equal to or greater than coefficients of unknown system to be identified. Otherwise accuracy drastically decreases. So estimation of number of coefficients in unknown plant is an important task performed by VL-LMS algorithms in the adaptive filter applications.

**3. IMPLEMENTATION METHODOLOGY**

The adopted methodology is Simulation of system identification application using Matlab. In the current experimentation model, two tap unknown system is selected. If the number of taps in adaptive filter is less than unknown system taps (here two) the adaptive filter is called as under specified; if more than that, then called as over specified, otherwise called as rightly specified. The under specified filter cannot satisfactorily perform system identification. In the current context we have selected rightly specified case and taken five coefficients in the filter weight.

The data points in training sequence are 1000. This program calculates the results for different values of μ.
The unknown system to be identified by the adaptive filter is taken as below.
h = [1 2];
The program runs for four different values of mu given as,
mu = [0.001 0.01 0.1 0.5];
The no of iterations is obtained at run time from user by following command.
iterations = input('Enter no of iterations');
Input signal is synthesized for desired no of iterations by following command.
x = randn(1,iterations);
The input signal x(n) is applied to both the unknown system and adaptive filter.
The desired signal d(n)is obtained by filtering the input signal with the unknown system function using following command.
d = conv(x,h);
Initial weight of adaptive filter are made equal to zero by,
w0(1:length(mu),1) = 0; w1(1:length(mu),1) = 0;
Filtering and adaptation takes place in the following loops:

```
for i = 1:length(mu)
y(1) = w0(i,1)*x(1);  % first iteration at n=0;
e(1) = d(1)- y(1);  % first error
w0(i,2) = w0(i,1)+mu(i)*e(1)*x(1);
w1(i,2) = w1(i,1)+mu(i)*e(1)*x(1);
   for n = 2:iterations   % lms loop
   y(n) = w0(i,n)*x(n)+w1(i,n)*x(n-1);
   e(n) = d(n) - y(n);
   w0(i,n+1) = w0(i,n)+mu(i)*e(n)*x(n);
   w1(i,n+1) = w1(i,n)+mu(i)*e(n)*x(n-1);
   end
  E(i,:) = e;
  cost(i,:) = e.*e;
end
mse_error = mse(e);
```

Additional program code is written to plot the important parameters.

## 4. RESULTS AND DISCUSSION

The simulation result in Fig. 5 shows the progress of iterated unknown plant coefficient towards the actual coefficient w.r.t. iterations for different values of μ. This figure indicates that for smaller values of μ, convergence is very slow and for larger values of mu, convergence is fast and initial transients are very high. From the results it is observed that initially there is larger difference in actual and estimated output, but as time progresses, estimated output goes closer and closer to the actual output.

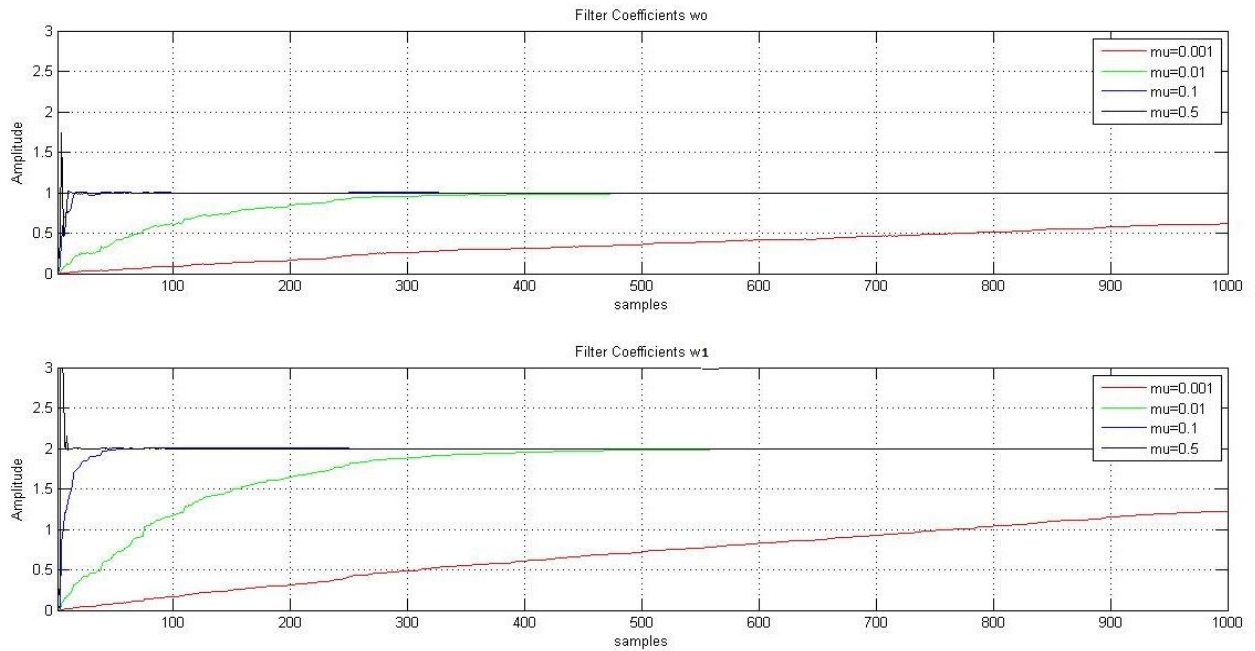

Fig.5. Filter Weight Update vs. iterations for different values of μ

Mean square error is obtained by using the Matlab function as given below.

MSE_error = mse(e).

Execution time is obtained using Matlab function tic –toc. tic starts the timer and toc stops the timer and displays the elapsed time.

The execution of program and command prompt input and output for 1000 iterations are given below.

Enter mu>>0.001

Enter no of iterations>>1000

Elapsed time is 0.008446 seconds.

MSE = 2.135379e+000

Similarly fig. 6 (a) shows instantaneous squared estimation error vs. iteration for 5000 iterations. Fig. 6 (b-i) shows the desired signal, output signal and error signal for 1000 iterations. Similarly, Fig 6(b-ii) shows actual system coefficients versus estimated system coefficients.

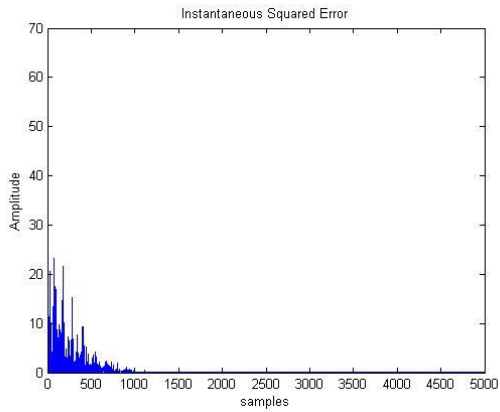
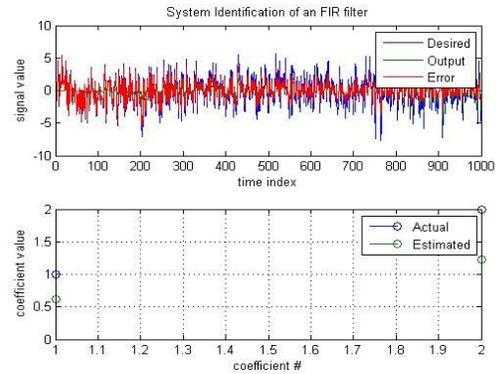

a) Squared Error vs. Iteration for 5000 iterations

b) Desired, Output, Error signal and Actual and Estimated weight

Fig.6. Filter Response

The effect of step size and no of iterations is evident from few more runs of the program given below. Here it is observed that the execution time increases and the MSE decrease because of increase in no of iterations. Here the step size is consistent with convergence rate parameter given in (20).

Enter mu>>0.002
Enter no of iterations>>5000
Elapsed time is 0.091981 seconds.
MSE = 2.510136e-001

In next run, even in 1000 iterations, the MSE is low because of higher step size.
Enter mu>>0.01
Enter no of iterations>>1000
Elapsed time is 0.015189 seconds.
MSE = 2.517429e-001

The following table calculates various parameters for different values of μ and different number of iteration. From this table it is observed that with step size of 0.001, 4770 iterations are required for system identification with over 99% accuracy. Similarly, with step size of 0.002, 2486 iterations are required. For further increase in step size, the number of required iterations continuously decreases, but after certain limit when step size is higher than 0.4, filter tends to be unstable and may increase MSE, and error in estimation.

Table1. System Identification Result for Different Values of μ

| Program: LMS_sys_ident_11.m | | | | | | | | | | | | |
|---|---|---|---|---|---|---|---|---|---|---|---|---|
| % SYSTEM IDENTIFICATION USING LMS ALGORITHM | | | | | | | | | | | | |
| Step Size and iteration optimization for accuracy over 99% | | | | | | | | | | | | |
| Mu | Iterations | MSE | h(1) | w0 | h(2) | w1 | Squared Error | Iteration for w0 | Iteration for w1 | Combined iteration | Accuracy in 1000 iterations | Loop Execution time (in sec) |
| 0.001 | 1000 | 2.23 | 1 | 0.7027 | 2 | 1.3269 | 0.465 | 4770 | 4659 | 4770 | 68.3075 | 0.087866 |
| 0.002 | 1000 | 1.23 | 1 | 0.8898 | 2 | 1.7463 | 0.069 | 2486 | 2453 | 2486 | 88.1475 | 0.028585 |
| 0.003 | 1000 | 0.83 | 1 | 0.9349 | 2 | 1.8765 | 0.015 | 1514 | 1512 | 1514 | 93.6575 | 0.019515 |
| 0.004 | 1000 | 0.63 | 1 | 0.9854 | 2 | 1.9648 | 0.001 | 951 | 1109 | 1109 | 98.39 | 0.018786 |

| 0.005 | 1000 | 0.50 | 1 | 0.9953 | 2 | 1.9887 | 0.000 | 824 | 872 | 872 | 99.4825 | 0.007870 |
| 0.006 | 1000 | 0.42 | 1 | 0.999 | 2 | 1.9978 | 0.000 | 801 | 807 | 807 | 99.895 | 0.011098 |
| 0.007 | 1000 | 0.36 | 1 | 0.9985 | 2 | 1.9976 | 0.000 | 622 | 648 | 648 | 99.865 | 0.006637 |
| 0.008 | 1000 | 0.32 | 1 | 0.9996 | 2 | 1.9993 | 0.000 | 594 | 542 | 594 | 99.9625 | 0.009648 |
| 0.009 | 1000 | 0.28 | 1 | 0.9998 | 2 | 1.9997 | 0.000 | 563 | 518 | 563 | 99.9825 | 0.009644 |
| 0.01 | 1000 | 0.25 | 1 | 1 | 2 | 1.9999 | 0.000 | 485 | 450 | 485 | 99.9975 | 0.006816 |
| 0.02 | 1000 | 0.13 | 1 | 1 | 2 | 2 | 0.000 | 252 | 242 | 252 | 100 | 0.005589 |
| 0.03 | 1000 | 0.09 | 1 | 1 | 2 | 2 | 0.000 | 159 | 137 | 159 | 100 | 0.002169 |
| 0.04 | 1000 | 0.07 | 1 | 1 | 2 | 2 | 0.000 | 117 | 122 | 122 | 100 | 0.004215 |
| 0.05 | 1000 | 0.05 | 1 | 1 | 2 | 2 | 0.000 | 87 | 91 | 91 | 100 | 0.003978 |
| 0.06 | 1000 | 0.04 | 1 | 1 | 2 | 2 | 0.000 | 85 | 68 | 85 | 100 | 0.003953 |
| 0.07 | 1000 | 0.037 | 1 | 1 | 2 | 2 | 0.000 | 50 | 60 | 60 | 100 | 0.003650 |
| 0.08 | 1000 | 0.030 | 1 | 1 | 2 | 2 | 0.000 | 29 | 43 | 43 | 100 | 0.003461 |
| 0.09 | 1000 | 0.030 | 1 | 1 | 2 | 2 | 0.000 | 40 | 46 | 46 | 100 | 0.003493 |
| 0.1 | 1000 | 0.027 | 1 | 1 | 2 | 2 | 0.000 | 43 | 34 | 43 | 100 | 0.003463 |
| 0.2 | 1000 | 0.018 | 1 | 1 | 2 | 2 | 0.000 | 5 | 11 | 11 | 100 | 0.002307 |
| 0.3 | 1000 | 0.009 | 1 | 1 | 2 | 2 | 0.000 | 5 | 6 | 6 | 100 | 0.001807 |
| 0.4 | 1000 | 0.018 | 1 | 1 | 2 | 2 | 0.000 | 2 | 3 | 3 | 100 | 0.003823 |
| 0.5 | 1000 | 0.009 | 1 | 1 | 2 | 2 | 0.000 | 20 | 14 | 20 | 100 | 0.003487 |
| 0.6 | 1000 | 0.024 | 1 | 1 | 2 | 2 | 0.000 | 2 | 2 | 2 | 100 | 0.001550 |
| 0.7 | 1000 | 0.003 | 1 | 1 | 2 | 2 | 0.000 | 12 | 13 | 13 | 100 | 0.003221 |
| 0.8 | 1000 | 0.010 | 1 | 1 | 2 | 2 | 0.000 | 2 | 4 | 4 | 100 | 0.003317 |
| 0.9 | 1000 | 64.53 | 1 | 1 | 2 | 2 | 0.000 | 4 | 4 | 4 | 100 | 0.003349 |
| 1 | 1000 | 7.3E9 | 1 | 4.6E+3 | 2 | 1.04E+04 | 1.1E+8 | 2 | 3 | 3 | NA | NA |

## 5. CONCLUSIONS

The experimental result indicates that the adaptive filter successfully estimates the unknown system coefficients and converges. We have successfully studied the effect of step size and number of iterations on the filter performance like means square error, estimation accuracy etc. The decrease in step size decreases steady state error but at the same time it increases the convergence time. Thus there is a trade-off between these two quantities.

### 5.1 Future Improvements:

After simulation study presented here, future improvements may be the hardware implementation of the designed algorithm.